\documentclass[ twocolumn, seceq ]{jpsj3}
\usepackage{graphicx}
\usepackage{dcolumn}
\usepackage{bm}
\def\runauthor{Fusayoshi J. \textsc{Ohkawa}}
\begin{document}
\title{%
Itinerant-Electron Magnetism in the Heisenberg Limit%
%
}
\author{Fusayoshi J. \textsc{Ohkawa}\thanks{E-mail address: fohkawa@mail.sci.hokudai.ac.jp} }
%
\inst{%
Department of Physics, Faculty of Science, 
Hokkaido University, Sapporo 060-0810, Japan}
\recdate{%
July 5, 2010}
%
%
\abst{
The Hubbard model in the Heisenberg limit is studied by Kondo-lattice theory.
The Kondo temperature $T_{\rm K}$ or $k_{\rm B}T_{\rm K}$, which is an energy scale of low-energy local quantum spin fluctuations,
is enhanced by the resonating valence bond (RVB) mechanism,
so that $T_{\rm K}\simeq T_{\rm MF}/(2D)$, where $T_{\rm MF}$ is the N\'{e}el temperature in the mean-field approximation of the corresponding Heisenberg model and $D$ is the spatial dimensionality.
Electrons certainly behave as localized spins at $T\gg T_{\rm K}$, but they are still itinerant at $T\ll T_{\rm K}$ unless an antiferromagnetic complete gap opens.
When the N\'{e}el temperature $T_{\rm N}$ is so high that $T_{\rm N}\gg T_{\rm MF}/(2D)$, magnetism is prototypic local-moment magnetism.
When $T_{\rm N}$ is so low that $T_{\rm N} \ll T_{\rm MF}/(2D)$ because of low dimensionality or frustration, magnetism 
is itinerant-electron magnetism of an almost spin liquid, i.e., a normal Fermi liquid or a Tomonaga-Luttinger liquid in which
the spectral weight of single-particle excitations is almost vanishing.
The spin susceptibility has a temperature and wave-number dependence characteristic of itinerant-electron magnetism.  
This type of itinerant-electron magnetism must also be possible in the Heisenberg model.
}
\kword{Hubbard model, Heisenberg model, itinerant-electron magnetism, local-moment magnetism, Kondo lattice, resonating valence bond, Fermi liquid, Tomonaga-Luttinger liquid, spin liquid, Mott insulator}
%
 %
\maketitle
\section{Introduction}
\label{SecIntroduction}
The problem of moment formation or magnetism is one of the most important issues in condensed-matter physics.\cite{revMag1,revMag2}
There are two types of magnetism, local-moment magnetism and itinerant-electron magnetism.
According to this nature of magnetism, there are two types of effective Hamiltonians, a localized-spin model such as the Heisenberg model and an itinerant-electron model such as the Hubbard model.
It is really a long-standing issue how to construct a unified theory that can explain the two types of magnetism.
When electrons are localized, an itinerant-electron model is reduced to a localized-spin model. 
First of all, the mechanism of electron localization should be elucidated.

The theory or scenario for electron localization proposed by Mott and other people,\cite{mott} such as the Mott insulator or the Mott-Hubbard metal-insulator transition, is controversial.
Consider the Hubbard model with the on-site repulsion $U$ and, e.g., only the nearest-neighbor transfer integral $t$.
When $U/|t|=+\infty$ and $N=L$,
where $N$ is the number of electrons and $L$ is that of unit cells, all the unit cells are singly occupied and there is no empty or double occupancy, so that no electron can be itinerant.
This ground state is a prototypic Mott insulator. 
It is a highly degenerate state such that it has an electron of arbitrary spin at each unit cell or the entropy is $k_{\rm B}\ln2$ per unit cell.  The third law of thermodynamics is broken in the prototypic Mott insulator.  
Then, one may suspect that, if the ground state is a Mott insulator for a finite $U$, the third law must also be broken in it; and the other may suspect that, since the third law must hold, the scenario itself is doubtful.

In the Heisenberg limit, the Hubbard model 
is reduced to the Heisenberg model with the superexchange interaction $J=-4t^2/U$.
When $J=0$, the ground state is certainly the prototypic Mott insulator.
Fazekas and Anderson\cite{fazekas} proposed a stabilization mechanism of a spin liquid by the formation of a local but itinerant singlet or a resonating valence bond (RVB) on each pair of nearest neighbors;
this spin liquid is called an RVB spin liquid and this stabilization mechanism is called an RVB mechanism.
When $J<0$, the prototypic Mott insulator is unstable against an RVB spin liquid. 

In order to avoid confusion, we tentatively restrict our discussion to the ground state within the constrained Hilbert subspace where no order parameter exists.
Either of a Mott insulator and a spin liquid is an insulator. They should be distinguished from each other.
In this paper, only an insulator where the third law is broken is called a Mott insulator. If the third law holds in an insulator, it is Wilson's insulator or a spin liquid according to the spectrum of low-energy spin excitations. 

According to Hubbard's theory,\cite{hubbard1,hubbard3} the Hubbard gap opens and the band splits into the upper and lower Hubbard bands.
When $U>2z|t|$ at least, where $z$ is the coordination number of nearest neighbors, the Hubbard gap is a complete gap and the ground state is a Mott insulator. 
However, the virtual process allowing empty and double occupancies is possible, unless $|t|/U=0$.
The superexchange interaction arises from the virtual process.\cite{Js-mech-pert} 
%
Since the Fock-type term of the superexchange interaction stabilizes a local singlet on each pair of nearest neighbors, it eventually stabilizes a singlet ground state or a normal Fermi liquid \cite{ohkawa-RVB} rather than a Mott insulator; this stabilization mechanism is simply the RVB mechanism. 
In a previous paper,\cite{toyama} it was proved based on Kondo-lattice theory \cite{Mapping-1,Mapping-2,Mapping-3} that
the ground state cannot be a Mott insulator,
unless $U/|t|=+\infty$ and $N=L$.
When $U$ is finite, the Hubbard gap is a pseudo-gap.

Hubbard's theory is within the single-site approximation (SSA).
If all the single-site terms are considered in a theory,
it is simply the supreme SSA (S$^3$A) theory.
The S$^3$A is rigorous for infinite dimensions,\cite{Metzner,Muller-H1,Muller-H2,Janis} but within the constrained Hilbert subspace. 
According to the Kondo-lattice theory,\cite{Mapping-1,Mapping-2,Mapping-3}
the S$^3$A is mapped to a problem of self-consistently determining and solving the Anderson model,\cite{Mapping-1,Mapping-2,Mapping-3}
which is an effective Hamiltonians for the Kondo effect.
The S$^3$A can also be formulated as the dynamical mean-field theory (DMFT)\cite{georges,PhyToday,RevMod,kotliar} or the dynamical coherent potential approximation (DCPA).\cite{dcpa}
It was also proved in the previous paper\cite{toyama} that, in the S$^3$A, i.e., even if the RVB mechanism is not considered, the ground state is a normal Fermi liquid.
The density of state has a three-peak structure, a quasi-particle band  between the upper and lower Hubbard bands, which corresponds to the three-peak structure in the Anderson model, the Kondo peak between two sub-peaks. The quasi-particle band is simply what is predicted by a combination of Gutzwiller's theory,\cite{Gutzwiller1,Gutzwiller2,Gutzwiller3} which is also within the SSA, and the Fermi-liquid theory.\cite{Luttinger1,Luttinger2}

One may suspect that the superexchange interaction cannot work in a metal since the formation of localized spins is assumed in the well-known derivation of the superexchange interaction.\cite{Js-mech-pert}
In field theory, the superexchange interaction arises from the virtual exchange of a pair excitation of an electron and a hole in the lower and upper Hubbard bands.\cite{sp1,sp2,sp3}
The superexchange interaction works in not only a spin liquid but also a metal where the Hubbard gap is a pseudo-gap. 

Multi-site terms are perturbatively considered by the Kondo-lattice theory.\cite{toyama,Mapping-1,Mapping-2,Mapping-3}
A normal Fermi liquid in the S$^3$A, which is stabilized by the Kondo effect, is further stabilized by the RVB mechanism.
The stabilized Fermi liquid is a normal Fermi liquid, which is totally frustrated.\cite{comFrustration} 
The Fermi liquid is a relevant {\it unperturbed} state to study a normal or anomalous Fermi liquid in the constrained or whole Hilbert space and an eventual true ground state in the whole Hilbert space.
The Kondo temperature $T_{\rm K}$ is an energy scale of low-energy local quantum spin fluctuations in a normal or anomalous Fermi liquid or an ordered state.   
Since the entropy is about $k_{\rm B}\ln 2$ per unit cell at $T\gg T_{\rm K}$, a high temperature phase at $T\gg T_{\rm K}$ is simply a Mott insulator. Electrons behaves as localized spins at $T \gg T_{\rm K}$.
The Kondo-lattice theory can treat not only itinerant electrons at $T\ll T_{\rm K}$ but also localized spins  at $T \gg T_{\rm K}$. 
It is interesting to study the Heisenberg limit of the Hubbard model by the Kondo-lattice theory in order to elucidate the mechanism of moment formation 
and the nature of itinerant-electron magnetism and local-moment magnetism.

One of the purposes of this paper is to show that the Kondo-lattice theory based on the Hubbard model can explain local-moment magnetism in the Heisenberg model.
Then, it will be shown that, in lower dimensions, 
itinerant-electron magnetism is possible even in the Heisenberg limit.
%
This paper is organized as follows.
Preliminary is given in \S\ref{SecPreliminary}.
The Heisenberg limit is studied in \S\ref{SecHeisenbergLimit}.
Discussion is given in \S\ref{SecDiscussion}.
Conclusion is given in \S\ref{SecConclusion}.
An equality is proved in Appendix\ref{SecEquality}.

\section{Preliminary}
\label{SecPreliminary}
\subsection{Two effective Hamiltonians}
\subsubsection{Hubbard model with an electron reservoir}
In the grand canonical ensemble, the averaged electron number is irrational, in general, and is a continuous function of the chemical potential. An electron reservoir should be explicitly considered to ensure this crucial property. 
In this paper, for simplicity, we assume that both the Hubbard model and the reservoir are on a hypercubic lattice in $D$ dimensions.

The total Hamiltonian is composed of four terms:
\begin{equation}\label{EqTotalHamil}
{\cal H} = {\cal H}_d + {\cal H}_c + {\cal V}_{d\mbox{-}c}
-\mu \left({\cal N}_d+{\cal N}_c\right).
\end{equation}
The first term is the Hubbard model:
\begin{equation}\label{EqHubbardH}
{\cal H}_d = 
\epsilon_d \sum_{i\sigma} n_{i\sigma}
- \frac{t_d}{\sqrt{D}} \sum_{\left<ij\right>\sigma} 
d_{i\sigma}^\dag d_{j\sigma}
+U \sum_{i} n_{i\uparrow}n_{i\downarrow} ,
\end{equation}
where $\epsilon_d$ is the band center, $d_{i\sigma}^\dag$ and $ d_{i\sigma}$ are creation and annihilation operators of an electron with spin $\sigma$ at the $i$th unit cell, $n_{i\sigma}=d_{i\sigma}^\dag d_{i\sigma}$, $-t_d/\sqrt{D}$ is the transfer integral between nearest neighbors $\left<ij\right>$, and $U$ is the on-site repulsion.
For convenience, the transfer integral  
includes dimensionality $D$,\cite{Metzner,Muller-H1,Muller-H2,Janis} so that an effective bandwidth is $O(|t_d|)$.
The second term stands for the reservoir:
\begin{equation}\label{EqRes}
{\cal H}_c = 
\epsilon_c\sum_{i\sigma}c_{i\sigma}^\dag c_{i\sigma}
- \frac{t_c}{\sqrt{D}} \sum_{\left<ij\right>\sigma} c_{i\sigma}^\dag c_{j\sigma} ,
\end{equation}
where every notation is conventional as in Eq.~(\ref{EqHubbardH}).
The third term is an infinitesimally small hybridization between the Hubbard model and the reservoir:
\begin{align}\label{EqReservoirV}
{\cal V}_{d\mbox{-}c} &= 
\lambda \sum_{i\sigma}
\left[
v_{i} d_{i\sigma}^\dag c_{i \sigma}
+v_{i}^* c_{i\sigma}^\dag d_{i \sigma} \right] ,
\end{align}
where $v_{i}$ is a hybridization matrix element at the $i$th unit cell; $v_{i}$ may be zero or nonzero at a unit cell. 
For convenience, an infinitesimally small but nonzero numerical constant, 
\begin{align}
\lambda= 0^+,
\end{align}
is introduced.
In the last term,  
\begin{align}
{\cal N}_d = \sum_{i\sigma} d_{i\sigma}^\dag d_{i\sigma},
\quad
{\cal N}_c = \sum_{i\sigma} c_{i\sigma}^\dag c_{i\sigma},
\end{align}
and $\mu$ is the chemical potential. 
The existence of the last term means that the Hubbard model and the reservoir are embedded in another electron reservoir.

Throughout of this paper, we assume that
\begin{align}\label{EqPH-Sym}
\mu= \epsilon_d + \frac1{2} U =  \epsilon_c .
\end{align}
Then,
$\bigl<{\cal N}_d\bigr>  = \bigl<{\cal N}_c\bigr>= L$, 
where $\left<\cdots\right>$ stands for the statistical average and $L$ is the number of unit cells. 
Only the just half-filling case is studied in this paper.
%
In order to recover the translational symmetry, for convenience, we consider averaged quantities over an ensemble of $v_{i}$'s, which is denoted as $\left\{v\right\}$, under an assumption such that $\left\{v\right\}$ is totally random, i.e.,
%
$\bigl<\hskip-3pt\bigl<\hskip1pt v_{i}\hskip1pt\bigr>\hskip-3pt\bigr> = 0$ and 
$\bigl<\hskip-3pt\bigl<\hskip1pt v_{i} v_{j}^*
\hskip1pt\bigr>\hskip-3pt\bigr> = \overline{|v|^2}\delta_{ij}$, 
%
where $\bigl<\hskip-3pt\bigl<\hskip1pt\cdots\hskip1pt\bigr>\hskip-3pt\bigr>$ stands for the ensemble average over $\left\{v\right\}$.
The periodic boundary condition is assumed.
The thermodynamic limit of $L\rightarrow +\infty$ is also assumed.

\subsubsection{Heisenberg model with a thermal reservoir}
\label{SecHeisenbergH}
The Heisenberg limit is defined by
$U/|t_d| \rightarrow +\infty$ with the constraint that 
\begin{align}\label{EqConstantJ}
J = - 4t_d^2/U ,
\end{align}
is kept constant,
together with the half-filling condition (\ref{EqPH-Sym}).
In this limit, the Hamiltonian (\ref{EqTotalHamil}) is reduced to 
\begin{align}
{\cal H}^\prime &= 
{\cal H}_d^\prime + {\cal H}_c + {\cal V}_{d\mbox{-}c}^{\hskip1pt\prime} -\mu{\cal N}_c + E_0.
\end{align}
If the Hilbert space is constrained to a subspace where  no empty or double occupancy exists such that
$n_{i\uparrow}+n_{i\downarrow} =1$,
\begin{align}
{\bm S}_i = \frac1{2}\sum_{\tau\tau^\prime}
{\bm \sigma}^{\tau\tau^\prime}
d_{i\tau}^\dag d_{i\tau^\prime},
\end{align}
where ${\bm \sigma}=(\sigma_x,\sigma_y,\sigma_z)$ is the Pauli matrix,
is simply a localized spin at the $i$th unit cell. 
The first term is the Heisenberg model:
\begin{align}\label{EqHeisenbergH}
{\cal H}_d^\prime &= 
-\frac1{2}\frac{J }{D}\sum_{\left<ij\right>} 
\left({\bm S}_i \cdot {\bm S}_j \right) .
\end{align}
The second term ${\cal H}_c$ is given by Eq.~(\ref{EqRes}).
The third term is an exchange interaction between the Heisenberg model and the reservoir:
\begin{align}
{\cal V}_{d\mbox{-}c}^{\hskip1pt\prime} &=
- \lambda^2  
\sum_{i}\sum_{\tau\tau^\prime}
K_{i}
\bigl({\bm S}_i \cdot {\bm \sigma}^{\tau\tau^\prime}\bigr)
c_{i\tau}^\dag c_{i\tau^\prime} ,
\end{align}
where $K_{i} = -2|v_{i}|^2/U$.
The interaction ${\cal V}_{d\mbox{-}c}^{\hskip1pt\prime}$ should be relevant in the Heisenberg limit, i.e.,
\begin{align}\label{EqJR}
\bigl<\hskip-3pt\bigl<\hskip0pt K_{i}
\hskip0pt\bigr>\hskip-3pt\bigr> 
&= - 2 \overline{|v|^2}/U ,
\end{align}
should be kept nonzero and finite. 
This condition should be included in the definition of the Heisenberg limit.
The fourth term is the chemical potential term for the reservoir. 
The last term  is a constant: $E_0=L(-U/2+J/2)$.

In the mean-field approximation, the static susceptibility of the Heisenberg model (\ref{EqHeisenbergH}) obeys the Curie-Weiss (CW) law for any wave number ${\bm q}$:
\begin{align}\label{EqChiMF}
\chi_s(0,{\bm q}) &=
1/\bigl[k_{\rm B}T - J_s({\bm q})/4\bigr],
\end{align}
where 
\begin{align}\label{EqSuperJs}
J_s({\bm q}) &=
\bigl(2J/\sqrt{D}\bigr)\varphi({\bm q}) ,
\end{align}
where
\begin{equation}\label{EqFormF}
\varphi({\bm q}) =  \frac1{\sqrt{D}}\sum_{\nu=1}^{D}\cos\left(q_{\nu}a\right),
\end{equation}
where $a$ is the lattice constant.
Throughout this paper, the susceptibility is defined in such a way that the conventional factor $g^2\mu_{\rm B}^2/4$ is not included, where $g$ is the $g$ factor and $\mu_{\rm B}$ is the Bohr magneton.
According to Eq.~(\ref{EqChiMF}), the N\'{e}el temperature is 
%
$T_{\rm MF} = 
-J/(2k_{\rm B})$,
%
and the ordering wave number of the N\'{e}el state is
${\bm Q}_{L}  = (\pm\pi/a, \pm\pi/a, \cdots, \pm\pi/a)$.
%
It will be shown in \S\ref{SecMagInfinite}
that this prototype of local-moment magnetism can be reproduced by the Kondo-lattice theory.

\subsection{Kondo-lattice theory}
\subsubsection{Kondo effect in the Hubbard model}
When $U=0$, the ensemble averaged single-particle Green function for electrons in the Hubbard model is given, in the wave number representation, by
\begin{align}
\bigl<\hskip-3pt\bigl<\hskip1pt
G_{\sigma}({\rm i}\varepsilon_n, {\bm k}) 
\hskip1pt\bigr>\hskip-3pt\bigr> &=
\frac1{{\rm i}\varepsilon_n +\mu-\epsilon_d -E({\bm k}) - \Gamma_d({\rm i}\varepsilon_n)},
\end{align}
where $\varepsilon_n=(2n+1)\pi k_{\rm B}T$, with $n$ being an integer, is a fermionic energy, 
\begin{align}
E({\bm k}) = - 2t_d \varphi({\bm k}),
\end{align}
with $\varphi({\bm k})$ defined by Eq.~(\ref{EqFormF}), and
\begin{align}
\Gamma_d({\rm i}\varepsilon_n) &=
\lambda^2 \overline{\left|v\right|^2} \frac1{L}\sum_{\bm k} 
\frac1{{\rm i}\varepsilon_n +\mu -\epsilon_c + 2t_c \varphi({\bm k})}.
\end{align}
In the following part of this paper, 
the notation $\bigl<\hskip-3pt\bigl<\hskip1pt \cdots\hskip1pt\bigr>\hskip-3pt\bigr>$ is not explicitly shown and, e.g., the ensemble averaged Green function is simply called the Green function.
Since $\lambda^2$ is infinitesimally small, terms of $O(\lambda^4)$ and higher are ignored in $\Gamma_d({\rm i}\varepsilon_n)$. It is obvious that
\begin{align}\label{EqNegativeImGammma}
{\rm Im }\Gamma_d(\varepsilon+{\rm i}0) & <0,
\end{align}
for $|\varepsilon|<2\sqrt{D}|t_c|$, where $\Gamma_d(\varepsilon+{\rm i}0)$ is the analytical continuation of $\Gamma_d({\rm i}\varepsilon_n)$. 
The Fermi surface is a hyper-surface composed of such ${\bm k}_{\rm F}$'s that satisfy
\begin{align}\label{EqFS}
\varphi({\bm k}_{\rm F})=0.
\end{align}
 
Within the Hilbert subspace where no order parameter exists, the Green function for nonzero $U$ is given by
\begin{align}\label{EqGreen}
G_{d\sigma}({\rm i}\varepsilon_n, {\bm k}) &=
1/\bigl[{\rm i}\varepsilon_n +\mu -\epsilon_d -E({\bm k}) 
\nonumber \\ &\qquad - 
 \Sigma_{\sigma}({\rm i}\varepsilon_n, {\bm k}) - \Gamma_d({\rm i}\varepsilon_n)\bigr],
\end{align}
where $\Sigma_{\sigma}({\rm i}\varepsilon_n, {\bm k})$ is the self-energy.
The self-energy is decomposed into the single-site $\tilde{\Sigma}_\sigma({\rm i}\varepsilon_n)$ and the multi-site $\Delta\Sigma_\sigma({\rm i}\varepsilon_n, {\bm k})$:
%
$\Sigma_\sigma({\rm i}\varepsilon_n, {\bm k}) =
\tilde{\Sigma}_\sigma({\rm i}\varepsilon_n)
+ \Delta\Sigma_\sigma({\rm i}\varepsilon_n, {\bm k})$.
%
To calculate the single-site $\tilde{\Sigma}_\sigma({\rm i}\varepsilon_n)$ is mapped to a problem of self-consistently determining the Anderson model and solving it.
In general, the Anderson model is characterized by the depth of the level of localized electrons measured from the chemical potential, the hybridization energy between the localized level and the conduction band, and the on-site repulsion.
The mapping condition is so simple that the depth and the on-site repulsion should be equal to $\epsilon_d-\mu$ and $U$, respectively, of the Hubbard model; and
the hybridization energy, which is denoted by $\Delta(\varepsilon)$, should be determined to satisfy 
\begin{align}\label{EqMap1}
\tilde{G}_{\sigma}({\rm i}\varepsilon_n) &=
R_{\sigma}({\rm i}\varepsilon_n),
\end{align}
where
\begin{align}\label{EqGreenAnderson}
\tilde{G}_{\sigma}({\rm i}\varepsilon_n) &=
\frac1{\displaystyle
{\rm i}\varepsilon_n+\mu  - \epsilon_d - \tilde{\Sigma}_\sigma({\rm i}\varepsilon_n)
- \frac1{\pi} \hskip-2pt \int\hskip-3pt d \varepsilon^\prime 
\frac{\Delta(\varepsilon^\prime)}
{{\rm i}\varepsilon_n-\varepsilon^\prime } },
\end{align}
is the Green function for the Anderson model and
\begin{align}\label{EqGreenR}
R_{\sigma}({\rm i}\varepsilon_n) &= 
(1/L)\sum_{\bm k}G_{d\sigma}({\rm i}\varepsilon_n,{\bm k}),
\end{align}
is the site-diagonal Green function for the Hubbard model.
Then, it follows from Eq.~(\ref{EqMap1}) that 
\begin{align}\label{EqMap2}
\Delta(\varepsilon) &=
 {\rm Im}\left[
\tilde{\Sigma}_\sigma(\varepsilon+{\rm i}0)
+1/R_{\sigma}(\varepsilon+{\rm i}0) \right] .
\end{align} 
As proved in the previous paper,\cite{toyama} in general,
\begin{align}\label{EqProofDelta}
\Delta(\varepsilon) \ge - {\rm Im }\Gamma_d(\varepsilon+{\rm i}0).
\end{align}
According to Eqs.~(\ref{EqNegativeImGammma}) and (\ref{EqProofDelta}), $\Delta(\varepsilon)>0$ for $|\varepsilon|<2|t_c|\sqrt{D}$.  
Since $\Delta(0)>0$ simply means that the Fermi surface exists in the conduction band of the Anderson model, it is proved that the ground state of the Anderson model is a normal Fermi liquid, even if the ground state of the Hubbard model is anomalous;
the single-site $\tilde{\Sigma}_\sigma({\rm i}\varepsilon_n)$ is exactly equal to the self-energy for the normal Fermi liquid in the Anderson model, even if the multi-site self-energy is anomalous.
\cite{toyama}  

Since the Anderson model is determined at each $T$, it includes $T$ as a parameter. The parameter $T$ is denoted as $T^\prime$ to distinguish it from $T$ itself.
In this paper,
the Kondo temperature $T_{\rm K}$, which is an energy scale of low-energy local quantum spin fluctuations, is defined by
%
\begin{align}\label{EqDefTK}
\bigl[\tilde{\chi}_s(0) \bigr]_{T=0\hskip2pt{\rm K}}&=
1/k_{\rm B} T_{\rm K}(T^\prime),
\end{align}
as a function of $T^\prime$,
where $\tilde{\chi}_s(0)$ is the susceptibility of the Anderson model determined at $T^\prime$, and $T=0$~K in the subscript means that the Anderson model is embedded in a reservoir of $T=0$~K.
Thus, $T_{\rm K}(T^\prime)$ or $T_{\rm K}(T)$ depends on $T$ in general.
If no confusion arises, $T_{\rm K}(T)$ is simply denoted as
$T_{\rm K}$ in this paper.
Since $\lambda^2> 0$,
$T_{\rm K}$ cannot be the absolute zero Kelvin, i.e., $T_{\rm K}>0$~K definitely.

The self-energy for the Anderson model in the presence infinitesimally small magnetic field $H$
is expanded in such a way that\cite{yamada1,yamada2}
\begin{align}\label{EqExpansionSigma}
\hskip-2pt
\tilde{\Sigma}_\sigma(\varepsilon+{\rm i}0) &=
\frac1{2} U + (1-\tilde{\phi}_\gamma) \varepsilon
+ \sigma (1-\tilde{\phi}_s)\frac1{2} g\mu_{\rm B} H
\nonumber \\ & \quad 
- {\rm i}\tilde{\phi}_{\rm w} \frac{\varepsilon^2+(\pi k_{\rm B}T)^2}{k_{\rm B}T_{\rm K}}+ \cdots,
\end{align}
for $|\varepsilon| \ll k_{\rm B}T_{\rm K}$ and $T\ll T_{\rm K}$, where $\tilde{\phi}_\gamma$, $\tilde{\phi}_s$, and $\tilde{\phi}_{\rm w}$ are positive constants.
%
Since charge fluctuations are almost completely suppressed in the Heisenberg limit, 
the Wilson ratio is two:\cite{yamada1,yamada2,wilson} 
\begin{align}
\tilde{W}_s = \tilde{\phi}_s/\tilde{\phi}_\gamma =2. 
\end{align}

According to the mapping condition (\ref{EqMap1}),
\begin{align}\label{EqRho}
\rho(\varepsilon) &=
- \frac1{\pi} {\rm Im}\tilde{G}_\sigma(\varepsilon+{\rm i}0) 
= - \frac1{\pi} {\rm Im}R_\sigma(\varepsilon+{\rm i}0).
\end{align}
%
Because of the particle-hole symmetry,
${\rm Re} \tilde{G}_\sigma(+{\rm i}0) ={\rm Re} R_\sigma(+{\rm i}0) =0$.
Then, it follows that\cite{comDMFT1}
%
\begin{align}\label{EqRhoDelta}
\pi \rho(0) \Delta(0) = 1.
\end{align}
If $\rho(0) \rightarrow 0$ or the ground state of the Hubbard model is approaching an insulating or spin-liquid phase,
 $\Delta(0)$ of the mapped Anderson model is diverging according to Eq.~(\ref{EqRhoDelta}); this property is crucial in \S\ref{SecFS} of this paper.

When the expansion (\ref{EqExpansionSigma}) is used, the Green function (\ref{EqGreen}) is written in such a way that
\begin{align}\label{EqCoherentG}
G_{d\sigma}^{(c)}(\varepsilon+{\rm i}0, {\bm k}) &=
\biggl[\tilde{\phi}_\gamma \varepsilon 
-E({\bm k}) 
+ {\rm i} \tilde{\phi}_{\rm w} \frac{\varepsilon^2+(\pi k_{\rm B}T)^2}{k_{\rm B}T_{\rm K}}
\nonumber \\ & \hskip-4pt
- \Delta\Sigma_\sigma(\varepsilon+{\rm i}0, {\bm k})
-\Gamma_d(\varepsilon+{\rm i}0)\biggr]^{-1} . 
\end{align}
This is accurate at $|\varepsilon| \ll k_{\rm B}T_{\rm K}$ and $T\ll T_{\rm K}$ and can be approximately used at $|\varepsilon| \lesssim k_{\rm B}T_{\rm K}$ and $T\lesssim T_{\rm K}$. 
The Green function (\ref{EqCoherentG}) is called a coherent part of the Green function; it describes only single-particle excitations in the vicinity of the chemical potential.

Because of ${\rm i} \tilde{\phi}_{\rm w}\varepsilon^2/k_{\rm B}T_{\rm K}$, no complete gap can open even if 
$\Delta\Sigma_\sigma(\varepsilon+{\rm i}0, {\bm k})$ diverges for any ${\bm k}$ as $\varepsilon \rightarrow 0$.
The ground state cannot be a Mott insulator nor an insulator proposed by Lieb and Wu,\cite{Lieb-Wu} in which the opening of a complete gap is expected.

%

Hubbard's theory or the Hubbard approximation
is only relevant at high energies $|\varepsilon| \gg k_{\rm B}T_{\rm K}$,\cite{hubbard1,hubbard3} i.e., 
the Green function in the Hubbard approximation can describe only single-particle excitations in the upper or lower Hubbard band.
In this paper, such a Green function that is only relevant at $|\varepsilon| \gg k_{\rm B}T_{\rm K}$ or $|\varepsilon| \gtrsim k_{\rm B}T_{\rm K}$ is called an incoherent part of the Green function.

\subsubsection{Intersite exchange interaction}
\label{SecIntersiteExcInt}
The irreducible polarization function in spin channels $\pi_s({\rm i}\omega_l, {\bm q})$ is also decomposed into the single-site $\tilde{\pi}_s({\rm i}\omega_l)$ and the multi-site $\Delta\pi_s({\rm i}\omega_l, {\bm q})$ such that
$\pi_s({\rm i}\omega_l, {\bm q})=
\tilde{\pi}_s({\rm i}\omega_l)+\Delta\pi_s({\rm i}\omega_l, {\bm q})$,
where $\omega_l = 2l\pi k_{\rm B}T$, with $l$ being an integer, is a bosonic energy.
The susceptibilities of the Anderson and Hubbard models are given by 
\begin{align}\label{EqSusAnderson}
\tilde{\chi}_s({\rm i}\omega_l) &=
2\tilde{\pi}_s({\rm i}\omega_l)/\bigl[1- U\tilde{\pi}_s({\rm i}\omega_l)\bigr] ,
\end{align}
and
\begin{align}\label{EqSusHubbard}
\chi_s({\rm i}\omega_l, {\bm q}) &=
2\pi_s({\rm i}\omega_l, {\bm q})/\bigl[1- U\pi_s({\rm i}\omega_l, {\bm q})\bigr] ,
\end{align}
respectively.
Since either of $\tilde{\chi}_s({\rm i}\omega_l)$ and
$\chi_s({\rm i}\omega_l, {\bm q})$ is nonzero and finite,
%
$U\tilde{\pi}_s({\rm i}\omega_l) \rightarrow 1$ and
$U\pi_s({\rm i}\omega_l,{\bm q}) \rightarrow 1$ 
%
should be satisfied in the Heisenberg limit, so that
%
$\Delta\pi_s({\rm i}\omega_l,{\bm q}) = O(1/U^2)$.
%
Then, it follows that
%
\begin{align}\label{EqSusKondoLattice}
\chi_s({\rm i}\omega_l, {\bm q}) &=
\frac{\tilde{\chi}_s({\rm i}\omega_l)}
{1- (1/4)I_s({\rm i}\omega_l, {\bm q}) \tilde{\chi}_s({\rm i}\omega_l)},
\end{align}
where
\begin{align}\label{EqIs}
I_s({\rm i}\omega_l, {\bm q}) = 2U^2 \Delta\pi_s({\rm i}\omega_l, {\bm q}).
\end{align}
Here, terms of $O\left[1/U\tilde{\chi}_s({\rm i}\omega_l)\right]$ and higher are ignored, because they vanish in the Heisenberg limit.

Equation~(\ref{EqSusKondoLattice}) is in full accord with
a physical picture of a Kondo lattice such that local spin fluctuations at different unit cells interact with each other by an intersite exchange interaction. The intersite exchange interaction is simply $I_s({\rm i}\omega_l, {\bm q})$.
It is decomposed into three terms:
\begin{align}\label{EqThreeExch}
I_s({\rm i}\omega_l, {\bm q}) = J_s({\bm q}) + 
J_Q({\rm i}\omega_l, {\bm q}) - 4 \Lambda({\rm i}\omega_l, {\bm q}).
\end{align}
The first term $J_s({\bm q})$ is the superexchange interaction,\cite{sp1,sp2,sp3} which is defined by Eq.~(\ref{EqSuperJs}).
The second term $J_Q({\rm i}\omega_l, {\bm q})$ is an intersite exchange interaction that arises from the virtual exchange of a low-energy pair excitation of an electron and a hole.
When the three-point single-site reducible and irreducible vertex function in spin channels are denoted by $\tilde{\Lambda}_s({\rm i}\varepsilon_n+{\rm i}\omega_l,{\rm i}\varepsilon_n; {\rm i}\omega_l)$ and $\tilde{\lambda}_s({\rm i}\varepsilon_n+{\rm i}\omega_l,{\rm i}\varepsilon_n; {\rm i}\omega_l)$, respectively,
it follows that 
\begin{align}
\tilde{\Lambda}_s({\rm i}\varepsilon_n+{\rm i}\omega_l,{\rm i}\varepsilon_n;{\rm i}\omega_l) &=
\frac{\tilde{\lambda}_s({\rm i}\varepsilon_n+{\rm i}\omega_l,{\rm i}\varepsilon_n; {\rm i}\omega_l)}{1- U \tilde{\pi}_s({\rm i}\omega_l)}.
\end{align}
In the Heisenberg limit, 
\begin{align}\label{EqLambdaRedIrred}
\tilde{\lambda}_s({\rm i}\varepsilon_n+{\rm i}\omega_l,{\rm i}\varepsilon_n; {\rm i}\omega_l)&=
\frac{
2\tilde{\Lambda}_s({\rm i}\varepsilon_n+{\rm i}\varepsilon_n,{\rm i}\varepsilon_n;{\rm i}\omega_l)  }
{U\tilde{\chi}_s({\rm i}\omega_l)} .
\end{align}
According to the Ward relation,\cite{ward}
\begin{align}\label{EqWard1}
\tilde{\Lambda}_s({\rm i}\varepsilon_n,{\rm i}\varepsilon_n;0) &=
1- \frac{d \phantom{h_{\rm Z}}}{d h_{\rm Z}}
\left[\tilde{\Sigma}_\uparrow({\rm i}\varepsilon_n) 
- \tilde{\Sigma}_\downarrow({\rm i}\varepsilon_n) \right],
\end{align}
where $h_{\rm Z}=g\mu_{\rm B} H$.
Thus, it follows from Eqs.~(\ref{EqExpansionSigma}), (\ref{EqLambdaRedIrred}), and (\ref{EqWard1}) that
\begin{align}\label{EqWard2}
\tilde{\lambda}_s({\rm i}\varepsilon_n,{\rm i}\varepsilon_n;{\rm i}\omega_l)
= 2\tilde{\phi}_s/\bigl[U\tilde{\chi}_s({\rm i}\omega_l)\bigr] ,
\end{align}
for $|\varepsilon_n|\ll k_{\rm B}T_{\rm K}$.
When Eq.~(\ref{EqWard2}) is approximately used as the single-site vertex function
for $|\varepsilon_n|\lesssim k_{\rm B}T_{\rm K}$ and
$|\omega_l|\lesssim k_{\rm B}T_{\rm K}$,
$J_Q({\rm i}\omega_l, {\bm q})$ is calculated to be
\begin{align}\label{EqJQ}
\hskip-2pt
J_Q({\rm i}\omega_l, {\bm q}) &=
\bigl[4\tilde{W}_s^2/\tilde{\chi}_s^2({\rm i}\omega_l)\bigr] 
\hskip-2pt \left[
P({\rm i}\omega_l, {\bm q}) - P_0({\rm i}\omega_l)
\right], 
\end{align}
where $\tilde{W}_s=2$ is the Wilson ratio, 
\begin{align}\label{EqPolP}
P({\rm i}\omega_l, {\bm q}) &=
- \frac{k_{\rm B} T}{L} \hskip-3pt\sum_{n{\bm k} \sigma} \!
\tilde{\phi}_\gamma^2
G_{d\sigma}^{(c)}({\rm i}\varepsilon_n \!+\! {\rm i}\omega_l, {\bm k} \!+\! {\bm q})
G_{d\sigma}^{(c)}({\rm i}\varepsilon_n, {\bm k}) ,
\end{align}
and
\begin{align}\label{EqPolP0}
P_0({\rm i}\omega_l) &=
\frac1{L}\sum_{\bm q}P({\rm i}\omega_l, {\bm q}).
\end{align}
%
The single-site term is subtracted in Eq.~(\ref{EqJQ}) to avoid double counting; $G_{d\sigma}^{(c)}({\rm i}\varepsilon_n, {\bm k})$ is the coherent part of the Green function; and, 
for convenience, $P({\rm i}\omega_l, {\bm q})$ is defined in such a way that it includes $\tilde{\phi}_\gamma^2$.
The third term $\Lambda({\rm i}\omega_l,{\bm q})$ is the sum of the whole remaining terms; it corresponds to the mode-mode coupling term 
in the self-consistent renormalization (SCR) theory of spin fluctuations.\cite{SCR}

When Eq.~(\ref{EqWard2}) is used, 
an exchange interaction mediated by intersite spin fluctuations is calculated to be
\begin{align}\label{EqSPinMedInt1}
\tilde{\lambda}_s^2(0,0;{\rm i}\omega_l)  U^2 \bigl[
\chi_s({\rm i}\omega_l,{\bm q}) - \tilde{\chi}_s({\rm i}\omega_l)
\bigr] &=
\tilde{\phi}_s^2 I_s^*({\rm i}\omega_l,{\bm q}),
\end{align}
where the single-site term is subtracted to avoid double counting and
\begin{align}\label{EqExchI*}
I_s^*({\rm i}\omega_l,{\bm q}) &=
\frac{I_s({\rm i}\omega_l,{\bm q})}
{1- (1/4)I_s({\rm i}\omega_l,{\bm q}) \tilde{\chi}_s({\rm i}\omega_l) } .
\end{align}
A perturbative theory in terms of $I_s({\rm i}\omega_l,{\bm q})$ can  treat the exchange interaction (\ref{EqSPinMedInt1});
$I_s({\rm i}\omega_l,{\bm q})$ is enhanced into $I_s^*({\rm i}\omega_l,{\bm q})$ by intersite spin fluctuations and, according to Eq.~(\ref{EqSPinMedInt1}), $\tilde{\phi}_s$ is an effective vertex function.
The perturbative theory is simply the Kondo-lattice theory and is also a $1/D$ expansion theory.

\section{Physical Properties of the Heisenberg Limit}
\label{SecHeisenbergLimit}
\subsection{Fermi surface of an RVB almost spin liquid}
\label{SecFS}
One of the most crucial terms in $I_s({\rm i}\omega_l,{\bm q})$ is the superexchange interaction $J_s({\bm q})$.
The first order term in $J_s({\bm q})$ is the Fock-type term or the RVB term:
\begin{align}\label{EqFock0}
\Delta\Sigma_\sigma^{\rm (RVB)}({\bm k}) &=
\frac{k_{\rm B}T}{L}\sum_{n{\bm p}\sigma^\prime}
\tilde{\phi}_s^2 \frac{1}{4}J_s({\bm k}-{\bm p})
\bigl({\bm \sigma}^{\sigma\sigma^\prime}
\hskip-3pt\cdot{\bm \sigma}^{\sigma^\prime\sigma}\bigr)
\nonumber \\ & \qquad \times
G_{d\sigma^\prime}^{\rm (c)}({\rm i}\varepsilon_n,{\bm p})e^{{\rm i}\varepsilon_n 0^+},
\end{align}
where $G_{d\sigma}^{(c)}({\rm i}\varepsilon_n,{\bm p})$ is 
defined by Eq.~(\ref{EqCoherentG}) and two $\tilde{\phi}_s$'s are included as two effective vertex functions.
When 
\begin{align}\label{EqJsKP}
J_s({\bm k} - {\bm p}) &=
\frac{2J}{D} \sum_{\nu=1}^{D}\bigl[
\cos(k_\nu a)\cos(p_\nu a)
\nonumber \\ & \qquad\qquad
+ \sin(k_\nu a)\sin(p_\nu a)\bigr],
\end{align}
is made us of, the RVB term is simply given by
\begin{align}\label{EqFock}
\Delta\Sigma_\sigma^{\rm (RVB)}({\bm k}) &=
(3/4) \tilde{\phi}_\gamma \tilde{W}_s^2 \Xi_{D}
(2J/D)\varphi({\bm k}),
\end{align}
where   
\begin{align}\label{EqXi}
\Xi_{D} &= 
\tilde{\phi}_\gamma\frac{k_{\rm B}T}{L} \sum_{n{\bm k}}
\varphi({\bm k})  G_{d\sigma}^{(c)}({\rm i}\varepsilon_n,{\bm k})e^{{\rm i}\varepsilon_n 0^+}.
\end{align}
For convenience, $\Xi_{D}$ is defined in such way that it includes $\tilde{\phi}_\gamma$.
When the RVB term is only considered in the multi-site self-energy,
$G_{d\sigma}^{(c)}(\varepsilon+{\rm i}o,{\bm k})$ is simply given by
\begin{align}\label{EqCoherentSimple}
G_{d\sigma}^{(c)}(\varepsilon+{\rm i}0,{\bm k}) &=
1/\bigl[\tilde{\phi}_\gamma\varepsilon + 2t_d^*\varphi({\bm k}) 
+ {\rm i}0\bigr],
\end{align}
where the life-time term proportional to $\tilde{\phi}_{\rm w}$ and $\Gamma_d(\varepsilon+{\rm i}0)$ are ignored and
\begin{align}\label{EqTd*}
t_d^* &=
t_d - (3/4)\tilde{\phi}_\gamma\tilde{W}_s^2
\Xi_{D} (J/D).
\end{align}
Then, it follows that 
\begin{align}\label{EqXi2}
\Xi_{D} &=
(1/L)\sum_{\bm k} \varphi({\bm k}) f\left[\xi({\bm k})\right],
\end{align}
where 
\begin{align}\label{EqLowerXi}
\xi({\bm k}) &=
-2(t_d^*/\tilde{\phi}_\gamma) \varphi({\bm k}),
\end{align}
and
\begin{align}
f(\varepsilon) =
1/\bigl[e^{\varepsilon/(k_{\rm B}T)}+1\bigr].
\end{align}
The density of state for the coherent part or for $|\varepsilon|\lesssim |t_d^*|$ is given by
\begin{align}\label{EqFLR0}
\rho^{\rm (c)}(\varepsilon) &=
\frac1{\tilde{\phi}_\gamma L}\sum_{\bm k}
\delta\bigl[\varepsilon- \xi({\bm k})\bigr] .
\end{align}
An effective bandwidth of $\rho^{\rm (c)}(\varepsilon)$ is about $|t_d^*|/\tilde{\phi}_\gamma$.
When $|\varepsilon|\lesssim |t_d^*|/\tilde{\phi}_\gamma$, 
\begin{align}\label{EqFLR1}
\rho^{\rm (c)}(\varepsilon) &=
\hskip-1pt\left\{\begin{array}{cl}
\displaystyle 
\hskip-5pt \bigl(\alpha_D/|t_d^*|\bigr)\hskip-2pt
\left[ 1 + O\bigl(\varepsilon^2\bigr)\right], 
& \hskip-5pt D\ne 2  \vspace{5pt} \\
\displaystyle
\hskip-5pt \bigl(\alpha_2/|t_d^*|\bigr) \hskip-1pt
\bigl[\ln\bigl| t_d^*/(\tilde{\phi}_\gamma\varepsilon)\bigr|
+ O\bigl(\varepsilon^0\bigr)\bigr], 
& \hskip-5pt D=2
\end{array} \right. \hskip-3pt ,
\end{align}
where $\alpha_D$ is a positive numerical constant; $\alpha_D=O(1)$.

The expansion coefficients $\tilde{\phi}_s$ and $\tilde{\phi}_\gamma$, which are single-site properties, should be consistently calculated with the RVB term.
In principle, this problem should be solved by self-consistently determining and solving the Anderson model. However, we take a different approach to this problem.
On the basis of Eq.~({\ref{EqRhoDelta}), we can assume that 
\begin{align}\label{EqFLR2}
\Delta(\varepsilon) &=
\left\{\begin{array}{cc}
|t_d^*|/(\pi\alpha_D) & |\varepsilon| \le |t_d^*|/(2\tilde{\phi}_\gamma) \\
0, & |\varepsilon| > |t_d^*|/(2\tilde{\phi}_\gamma)
\end{array} \right. ,
\end{align}
for  the Anderson model.
Since the probability of empty or double occupancy in the Hubbard model, which is approximately given by $(t_d^2/D)/U^2$ provided that the virtual processes allowing empty and  double occupancies are explicitly considered, i.e., the RVB mechanism is considered,\cite{comDMFT1,comDMFT2} should be equal to that in the Anderson model, which is approximately given by $\Delta(0)(|t_d^*|/\tilde{\phi}_\gamma)/ U^2$, an approximate relation such that
\begin{align}\label{EqFLR3}
t_d^2/\bigl(DU^2\bigr) &\simeq
\Delta(0)|t_d^*|\big/\bigl(\tilde{\phi}_\gamma U^2\bigr),
\end{align}
should hold.  It follows from Eqs.~(\ref{EqTd*}), (\ref{EqFLR2}), and (\ref{EqFLR3}) that
%
$\tilde{\phi}_\gamma \simeq
\alpha_D \pi U\big/\bigl[3^2 \hskip1pt \tilde{W}_s^2 \Xi_{D}^2 (J/D)\bigr]$.
%
Then, 
\begin{align}\label{EqFLR5}
\tilde{\phi}_s=2\tilde{\phi}_\gamma =O\bigl(DU/|J|\bigr),
\hskip5pt
|t_d^*| = O\bigl(U\bigr),
\end{align}
and
\begin{align}\label{EqFLR6}
\rho^{\rm (c)}(0)=1/[\pi\Delta(0)] = O\bigl(1/U\bigr).
\end{align}
In the Heisenberg limit, $\tilde{\phi}_\gamma$ and $|t_d^*|$ are  diverging and $\rho^{\rm (c)}(\varepsilon)$ is vanishing, so that the ground state within the constrained Hilbert subspace is an almost spin liquid.\cite{comDMFT1} 

When $D\ne 2$, $T_{\rm K}$ can be evaluated from the Fermi-liquid relation for the Anderson model:\cite{yamada1,yamada2}
\begin{align}\label{EqFLR7}
\tilde{\chi}_s(0) &= 
2\tilde{\phi}_s \rho^{\rm (c)}(0),
\end{align}
at $T=0$~K. 
It follows from Eqs.~(\ref{EqDefTK}) and (\ref{EqFLR7}) that
\begin{align}\label{EqFLR8}
\rho^{\rm (c)}(0)= 1/(2\tilde{\phi}_s k_{\rm B}T_{\rm K}).
\end{align}
It follows from Eqs.~(\ref{EqTd*}), (\ref{EqFLR1}), and (\ref{EqFLR8}) that
\begin{align}\label{EqFLR9}
k_{\rm B}T_{\rm K} &\simeq |t_d^*|/\bigl(4\tilde{\phi}_\gamma \bigr)
\simeq 3\Xi_{D}|J|/(4D) . 
\end{align}
When $T=0$~K, $\Xi_D$'s are calculated to be $\Xi_{1}=1/\pi=0.31831\cdots$, $\Xi_{2}=2\sqrt{2}/\pi^2=0.28658\cdots$, and $\Xi_{\infty}=1/(2\sqrt{\pi})=0.283095\cdots$, i.e., 
$\Xi_{D}\simeq 1/3$ for any $D$. When $D\ne 2$, it follows that
\begin{align}\label{EqTK-TMF}
T_{\rm K} &\simeq 
T_{\rm MF}/(2D),
\end{align}
where $T_{\rm MF}=|J|/(2k_{\rm B})$ is the N\'{e}el temperature in the mean-field approximation of the Heisenberg model.

When $D=2$, $\tilde{\chi}_s(0)$ has a logarithmic singularity due to that of $\rho^{\rm (c)}(\varepsilon)$.
The Anderson model is determined at each $T$ or $T^\prime$, as discussed above.
When the life-time term 
in $\tilde{\Sigma}_\sigma(\varepsilon+{\rm i}0)$ is considered in the mapping process at $T^\prime$, the singularity is suppressed, so that
\begin{align}
\bigl[\tilde{\chi}_s(0)
\bigr]_{T=0{\rm K}}
\simeq \bigl(2\alpha_2\tilde{\phi}_s/|t_d^*|\bigr)
\ln\bigl|(t_d^*/\tilde{\phi}_\gamma)/(
 k_{\rm B}T^\prime)\bigr|.
\end{align}
Then, $T_{\rm K}$ has a logarithmic $T$ dependence such that
\begin{align}\label{EqTK2D}
T_{\rm K} (T)&\simeq 
(T_{\rm MF}/4)\big/\ln\bigl|(t_d^*/\tilde{\phi}_\gamma)/
( k_{\rm B}T)\bigr|.
\end{align}
%
Unless $T$ is extremely low, Eq.~(\ref{EqTK-TMF}) is relevant even for $D=2$.

Note that $\Xi_{D}$ given by Eq.~(\ref{EqXi}) or (\ref{EqXi2}) is a decreasing function of $T$. Since $\Xi_{D}\rightarrow +0$ as $T\rightarrow +\infty$~K, $T_{\rm K}(T) \rightarrow +0$~K as $T\rightarrow +\infty$~K.
However, when $T\ll T_{\rm MF}/(2D)$, 
$T \ll T_{\rm K} (T)$ is satisfied for any $D$, even for $D=2$.

The multi-site self-energy $\Delta\Sigma_\sigma({\rm i}\varepsilon_n, {\bm k})$ is decomposed into the RVB term and the other term
$\delta\Sigma_\sigma({\rm i}\varepsilon_n, {\bm k})$:
\begin{align}
\Delta\Sigma_\sigma({\rm i}\varepsilon_n, {\bm k}) &=
\Delta\Sigma_\sigma^{\rm (RVB)}({\bm k})
+ \delta\Sigma_\sigma({\rm i}\varepsilon_n, {\bm k}) .
\end{align}
The RVB term does not depend on ${\rm i}\varepsilon_n$, so that it is quite normal.
If $\delta\Sigma_\sigma({\rm i}\varepsilon_n, {\bm k})$ is ignored,
therefore, the ground state is a normal Fermi liquid  
even in the Heisenberg limit. Although the Fermi liquid is an almost spin liquid,
the dispersion relation $\xi({\bm k})$ of quasi-particles is well defined and the Fermi surface can be unambiguously defined.
The Fermi surface is exactly the same as that for $U=0$, which is the hyper-surface defined by $\varphi({\bm k}_{\rm F}) =0$.

Since the perturbative scheme in terms of $I_s({\rm i}\omega_l,{\bm q})$ is in parallel to the conventional one in terms of $U$, it is straightforward to calculate 
$\delta\Sigma_\sigma({\rm i}\varepsilon_n, {\bm k})$.
According to a preliminary study, anomalous terms exist in $\delta\Sigma_\sigma({\rm i}\varepsilon_n, {\bm k})$ for $D=1$, 
which implies that the ground state is a Tomonaga-Luttinger (TL) liquid in one dimension.
If the Fermi surface is defined by $\varphi({\bm k}_{\rm F})=0$, it shows a perfect nesting for any $D$. Then, electrons may behave as a TL liquid in any $D$ dimensions, in particular, in two dimensions.\cite{andersonTM}
Even if the ground state is  a TL liquid, $\delta\Sigma_\sigma(\varepsilon+{\rm i}0, {\bm k})$ is continuos at $\varepsilon=0$ and its imaginary part vanishes at $\varepsilon=0$.
Then, $\delta\Sigma_\sigma(+{\rm i}0, {\bm k}_{\rm F})=0$ for any ${\bm k}_{\rm F}$ that satisfies $\varphi({\bm k}_{\rm F}) =0$ because of the particle-hole symmetry.
Within the preliminary study,
no anomalous term can be found for any $D$ such that the Fermi surface cannot be defined because of it.
If the Fermi surface can be defined, which is quite probable, it must be exactly the same as that for $U=0$.



\subsection{Metallic conductivity at $T=0$\hskip1pt{\rm K}}
\label{SecConductivity}
According to the Kubo formula,\cite{kubo} the conductivity is given by
\begin{align}
\sigma_{xx}(\omega) = 
(\hbar/{\rm i}\omega)\left[
K_{xx}(\omega+{\rm i}0)-K_{xx}(0)
\right],
\end{align}
where $K_{xx}(\omega+{\rm i}0)$ is the analytical continuation of 
\begin{align}\label{EqKXXT}
K_{xx}({\rm i}\omega_l) &= 
\frac1{La^D}\int_{0}^{\beta}  d\tau
e^{{\rm i}\omega_l\tau}\left< e^{\tau{\cal H}}
\hskip1pt \hat{j}_x \hskip1pt e^{-\tau{\cal H}} \hskip1pt \hat{j}_x\right> ,
\end{align}
where $\beta=1/(k_{\rm B}T)$ and $\hat{j}_x$ is
the $x$ component of the current operator defined by
\begin{align}
\hat{\bm j} &=
\sum_{{\bm k}\sigma} {\bm j}({\bm k}) n_{{\bm k}\sigma}, 
\end{align}
where 
\begin{align}
{\bm j}({\bm k}) &= 
-(e/\hbar)
(\partial/\partial{\bm k})
\bigl[ - 2t_d \varphi({\bm k}) \bigr],
\end{align}
and 
\begin{align}
n_{{\bm k}\sigma}= (1/L)\sum_{ij} e^{i{\bm k}\cdot\left({\bm R}_i-{\bm R}_j\right)} d_{i\sigma}^\dag d_{j\sigma}, 
\end{align}
where ${\bm R}_i$ is the position of the $i$th unit cell.
Equation~(\ref{EqKXXT}) is also described as 
\begin{align}
K_{xx}({\rm i}\omega_l) &= 
\frac{e^2}{\hbar^2}\frac{(2t_d)^2}{D a^{D-2}}
 \Pi_{xx}({\rm i}\omega_l),
\end{align}
where
\begin{align}
\Pi_{xx}({\rm i}\omega_l) &=
\frac1{L}\sum_{{\bm k}{\bm p}}\sin(k_x a)\sin(p_x a)
\int_{0}^{\beta} d\tau
e^{{\rm i}\omega_l\tau} 
\nonumber \\ & \quad \times 
\sum_{\sigma\sigma^\prime}
\left<e^{\tau{\cal H}}\hskip1pt n_{{\bm k}\sigma}\hskip1pt e^{-\tau{\cal H}}
\hskip1pt n_{{\bm p}\sigma^\prime}\right>.
\end{align}

First, we consider only the RVB term in the multi-site self-energy.
The current vertex ${\bm j}({\bm k})$ should be consistently renormalized with the RVB term to satisfy the Ward relation.\cite{ward}
The ladder vertex first order in $J_s({\bm q})$ corresponds to the RVB term, which is first order in $J_s({\bm q})$.
When Eq.~(\ref{EqJsKP}) is used, it is easy to show that 
\begin{align}\label{EqUpperPiJ}
\Pi_{xx}({\rm i}\omega_l)&=
\frac{1}{\tilde{\phi}_\gamma^2}
\frac{2\pi_{xx}({\rm i}\omega_l)}
{\displaystyle 1 + (3J/2D)\tilde{W}_s^2\pi_{xx}({\rm i}\omega_l)}, 
\end{align}
where 
\begin{align}\label{EqLowerPiJ}
\pi_{xx}({\rm i}\omega_l) &= 
- \tilde{\phi}_\gamma^2 \frac{k_{\rm B}T}{L}\sum_{n{\bm k}}\sin^2(k_x a)
G_{d\sigma}^{(c)}({\rm i}\varepsilon_n,{\bm k})
\nonumber \\ & \quad \times
G_{d\sigma}^{(c)}({\rm i}\varepsilon_n + {\rm i}\omega_l,{\bm k}).
\end{align}
%
For convenience, $\pi_{xx}({\rm i}\omega_l)$ is defined in such a way that it includes $\tilde{\phi}_\gamma^2$. 
When
\begin{align}\label{EqTwoVerJ1}
\frac{d \Pi_{xx}({\rm i}\omega_l)}
{d({\rm i}\omega_l)} &=
\frac{2}{\tilde{\phi}_\gamma^2}
\frac{\displaystyle 
d \pi_{xx}({\rm i}\omega_l)/d({\rm i}\omega_l)    }
{\displaystyle \left[1 + (3J/2D)\tilde{W}_s^2\pi_{xx}({\rm i}\omega_l)\right]^2},
\end{align}
is used, the static conductivity is calculated to be
\begin{align}\label{EqTwoVer2}
{\rm Re} \hskip1pt \sigma_{xx}(0) &=
\frac{e^2}{\hbar^2}\frac{(2t_d)^2}{Da^{D-2}}
\frac{S_{xx}(0)}{\displaystyle \left[1 + (3J/2D) \tilde{W}_s^2\pi_{xx}(0)\right]^2},
\end{align}
where
\begin{align}
S_{xx}(0) &=
\frac{2}{\tilde{\phi}_\gamma^2}
\lim_{\omega\rightarrow 0}
{\rm Re} \hskip2pt
\frac{\hbar}{{\rm i}\omega} \left[
\pi_{xx}(\omega+{\rm i}0) - \pi_{xx}(0)
\right].
\end{align}
When Eq.~(\ref{EqCoherentSimple}) is used for $G_{d\sigma}^{(c)}({\rm i}\varepsilon_n,{\bm k})$, it follows that
\begin{align}
\pi_{xx}(0) &=
\frac1{L} \sum_{\bm k}\sin^2(k_xa)
\left[-\frac{d f(\varepsilon)}{d\varepsilon}\right]_{\varepsilon=\xi({\bm k})},
\end{align}
%
where $\xi({\bm k})$ is defined by Eq.~(\ref{EqLowerXi}), and
\begin{align}\label{EqPJ1}
S_{xx}(0) &=
\frac{2\hbar}{\pi L} \sum_{\bm k}\sin^2(k_xa)
\int_{-\infty}^{+\infty} d\varepsilon 
\left[-\frac{f(\varepsilon)}{d \varepsilon }\right]
\nonumber \\ & \quad \times 
\left[{\rm Im} G_\sigma^{(c)}(\varepsilon+{\rm i}0,{\bm k})\right]^2.
\end{align}
When $T=0$~K, Eq.~(\ref{EqPJ1}) is reduced to
\begin{align}\label{EqPJ2}
S_{xx}(0) &=
\frac{2 \hbar}{\pi L}\hskip-2pt \sum_{\bm k}\sin^2(k_xa) \hskip-2pt
\left[{\rm Im} \frac1{-2t_d^* \varphi({\bm k}) - \Gamma_d(+{\rm i}0)}\hskip-1pt \right]^2 \hskip-1pt,
\end{align}
where the reservoir term $\Gamma_d(+{\rm i}0)$ is explicitly shown.
This is approximately calculated to be
\begin{align}
S_{xx}(0) &\simeq 
\frac{\hbar}{\pi |t_d^*|}\int_{-\infty}^{+\infty} d\varepsilon\left[ {\rm Im} \frac1{\varepsilon - {\rm i}\hskip2pt{\rm Im}\Gamma_d(+{\rm i}0)}\right]^2
\nonumber \\ &=
\frac{\hbar}{2\bigl|t_d^* \hskip2pt {\rm Im}\Gamma_d(+{\rm i}0)\bigr|}.
\end{align}
According to Eq.~(\ref{EqXiPi}) in Appendix\ref{SecEquality}, it follows that
\begin{align}\label{EqWardtdtd*}
t_d \left[1 + (3J/2D) \tilde{W}_s^2\pi_{xx}(0)\right]^{-1} &=
t_d^*.
\end{align}
Although $t_d=O\bigl(\sqrt{U}\bigr)$, $t_d^*=O\bigl(U\bigr)$. 
Then, it follows that
\begin{align}
{\rm Re}\sigma_{xx}(0) &\simeq
\frac{e^2}{2\hbar D a^{D-2}} \frac{|t_d^*|}{\bigl|{\rm Im}\Gamma_d(+{\rm i}0)\bigr|} .
\end{align}
Since $t_d^*=O(U)$ and ${\rm Im}\Gamma_d(+{\rm i}0)=O(\lambda^2U)$,
${\rm Re}\sigma_{xx}(0) = O\left(U^0/\lambda^2\right)$.
The conductivity diverges as $\lambda^2\rightarrow 0$.
Even if 
$\delta\Sigma_\sigma({\rm i}\varepsilon_n, {\bm k})$ is considered in addition to the RVB term,
this conclusion does not change, as studied below.

When the renormalized current vertex is denoted by
${\bm J}({\bm k};{\rm i}\varepsilon_n,{\rm i}\varepsilon_n+{\rm i}\omega_l)$, it follows that
%
\begin{align}
K_{xx}({\rm i}\omega_l) &=
-\frac{k_{\rm B}T}{La^D}\hskip-2pt\sum_{n{\bm k}\sigma}
j_x({\bm k})
G_\sigma^{(c)}({\rm i}\varepsilon_n,{\bm k})G_\sigma^{(c)}({\rm i}\varepsilon_n \!+\! {\rm i}\omega_l,{\bm k})
\nonumber \\ & \qquad \times
J_x({\bm k};{\rm i}\varepsilon_n,{\rm i}\varepsilon_n \!+\! {\rm i}\omega_l).
\end{align}
According to Eq.~(\ref{EqTwoVerJ1}), each $\pi_{xx}({\rm i}\omega_l)$ independently contributes to the ${\rm i}\omega_l$ linear term of $\Pi_{xx}({\rm i}\omega_l)$. According to a similar argument, 
as regards the ${\rm i}\omega_l$ linear term, it is sufficient to consider
\begin{align}
K_{xx}^{\prime}({\rm i}\omega_l) &=
-\frac{k_{\rm B}T}{La^D}\sum_{n{\bm k}\sigma}
G_\sigma^{(c)}({\rm i}\varepsilon_n,{\bm k})G_\sigma^{(c)}({\rm i}\varepsilon_n+{\rm i}\omega_l,{\bm k})
\nonumber \\ & \quad \times
J_x^2({\bm k};{\rm i}\varepsilon_n,{\rm i}\varepsilon_n),
\end{align}
instead of $K_{xx}({\rm i}\omega_l)$.
According to the Ward relation,\cite{ward}
\begin{align}\label{EqWardJ}
{\bm J}({\bm k}; {\rm i}\varepsilon_n, {\rm i}\varepsilon_n) = 
-\frac{e}{\hbar}
\frac{\partial\phantom{\bm k}}{\partial{\bm k}}
\Bigl[E_d({\bm k})
+ \Delta\Sigma_\sigma({\rm i}\varepsilon_n, {\bm k})
\Bigr].
\end{align}
The relation (\ref{EqWardtdtd*}) is consistent with Eq.~(\ref{EqWardJ}).
Although 
$\left|{\bm j}({\bm k})\right| = O\bigl(\sqrt{U}\bigr)$,
%
$\left|{\bm J}({\bm k}; {\rm i}\varepsilon_n, {\rm i}\varepsilon_n)\right|_{\varepsilon_n\rightarrow 0}=O(U)$.
%
If the current vertex is consistently renormalized with the multi-site self-energy, ${\rm Re}\sigma_{xx}(0) = O\left(U^0/\lambda^2\right)$. 
The conductivity at $T=0$~K is nonzero and is diverging as $\lambda^2\rightarrow 0$ even in the Heisenberg limit.\cite{abrahams}

\subsection{Two types of magnetism}
\label{SecMagnetism}
\subsubsection{Local-moment magnetism in higher dimensions}
\label{SecMagInfinite}
According to Eqs.~(\ref{EqSusKondoLattice}) and (\ref{EqThreeExch}), the inverse of the static susceptibility is given by
%
\begin{align}\label{EqInverseSus}
\frac1{\chi_s(0,{\bm q})} =
\frac1{\tilde{\chi}_s(0)}- \frac1{4}J_s({\bm q} )
- \frac1{4} J_Q(0,{\bm q}) + \Lambda(0,{\bm q}).
\end{align}
Then, we consider a function $T_{\rm N}({\bm q})$ defined by
\begin{align}\label{EqAFCondition}
\bigl[ 1/ \chi_s(0,{\bm q}) \bigr]_{T=T_{\rm N}({\bm q})}=0.
\end{align}
The N\'{e}el temperature $T_{\rm N}$ is simply the highest $T_{\rm N}({\bm q})$ as a function of ${\bm q}$; ${\bm q}$ that gives the highest $T_{\rm N}$ is an ordering wave number of the N\'{e}el state.

When Wilson's result for the $s$-$d$ model\cite{wilson} is extended to the $s$-$d$ limit of the Anderson model, it follows that
\begin{align}
\frac1{\tilde{\chi}_s(0)} &=
\left\{\begin{array}{cc}
k_{\rm B} T_{\rm K}, & T \ll T_{\rm K} \\
k_{\rm B} (T+ \theta_{\rm K}), & T \gg T_{\rm K}
\end{array}\right. ,
\end{align}
where $\theta_{\rm K}>0$ and $\theta_{\rm K}=O(T_{\rm K})$;
the definition of $T_{\rm K}$ in this paper is different by a numerical factor from Wilson's definition.
When $T\ll T_{\rm K}$, electrons {\it locally} behave as a normal Fermi liquid.
When $T\gg T_{\rm K}$,
electrons at each unit cell behave as a free localized spin.

First, we consider infinite dimensions $(D\rightarrow +\infty)$.
Since $T_{\rm K}\rightarrow +0$\hskip2ptK as $D\rightarrow +\infty$, 
$1/\tilde{\chi}_s(0) = k_{\rm B} T$.
Since $T/T_{\rm K}\rightarrow +\infty$ for any nonzero $T$, $J_Q(0,{\bm q})$ is vanishing for any ${\bm q}$.\cite{comItinerantMag} 
Since $D\rightarrow +\infty$, $\Lambda(0,{\bm q})$ is vanishing for any ${\bm q}$.
Although $J_s({\bm q})$ is $O\bigl(J/\sqrt{D}\bigr)$ and vanishing for almost all ${\bm q}$'s, it is $O(J/D^0)$ and nonzero for particular ${\bm q}$'s; e.g.,
\begin{align}
- J_s(0)= J_s({\bm Q}_{L})= - 2J =4k_{\rm B}T_{\rm MF}, 
\end{align}
where
${\bm Q}_{L} =
\left(\pm\pi/a, \pm\pi/a, \cdots, \pm\pi/a \right)$.
 Then, it is quite easy to show that
\begin{align}
T_{\rm N} = T_{\rm MF},
\end{align}
the ordering wave number of the N\'{e}el state is ${\bm Q}_L$, 
and the susceptibility at $T> T_{\rm MF}$ is exactly given by Eq.~(\ref{EqChiMF}).
In infinite dimensions, the susceptibility of the Heisenberg limit exactly obeys the CW law for prototypic local-moment magnetism.
Thus, magnetism in infinite dimensions is prototypic local-moment magnetism.

In finite dimensions,  
$T_{\rm K}\simeq T_{\rm MF}/(2D)$.
The reduction of $T_{\rm N}$ from $T_{\rm MF}$ is a higher order effect in $1/D$. When $D \gg 1$, therefore, it is obvious $T_{\rm MF} \gg T_{\rm K}$ and $T_{\rm N} \gg T_{\rm K}$. When $T>T_{\rm N}$, i.e., $T\gg T_{\rm K}$, $J_Q(0,{\bm q})$ is small for any ${\bm q}$. 
When $D\gg 1$, $\Lambda(0,{\bm q})$ is also small. 
At $T> T_{\rm N}$,  
the inverse of the susceptibility is approximately given by
\begin{align}\label{EqCW1}
\frac1{\chi_s(0,{\bm q}) } &\simeq 
k_{\rm B}(T + \theta_{\rm K}) 
- \frac1{4}J_s({\bm q} ) .
\end{align}
The N\'{e}el temperature is approximately given by 
\begin{align}\label{EqTNFinLoc}
T_{\rm N}\simeq
T_{\rm MF} -\theta_{\rm K} ,
\end{align}
where $\theta_{\rm K}=O\left[T_{\rm MF}/(2D)\right]$.
When $T_{\rm N}\gg T_{\rm K}$, magnetism is local-moment magnetism even in finite dimensions.

\subsubsection{Itinerant-electron magnetism in lower dimensions}
\label{SecItinerantM}
Because of critical fluctuations, no order appears at $T>0$~K in one and two dimensions.\cite{mermin} Then, $T_{\rm N}$ can be low in actual quasi-one and quasi-two dimensional magnets; $T_{\rm N}$ can also be low in actual highly frustrated magnets.
Here, we assume that $T_{\rm N} \ll T_{\rm K}$.
When $T_{\rm N}<T\ll T_{\rm K}$,
Eq.~(\ref{EqInverseSus}) is reduced to
\begin{align}\label{EqChiHighTK}
\frac1{\chi_s(0,{\bm q})} &=
k_{\rm B}T_{\rm K} - \frac1{4}J_s({\bm q})
-\frac1{4}J_Q(0,{\bm q})
+\Lambda(0,{\bm q}) . 
\end{align}
This can be approximately used at $T\lesssim T_{\rm K}$.
The $T$ dependence of $1/\chi_s(0,{\bm q})$ arises from 
$J_Q(0,{\bm q})$ and $\Lambda(0,{\bm q})$.

When $\delta\Sigma_\sigma({\rm i}\varepsilon_n, {\bm q})$ is ignored,
%
the density of states for quasi-particles is defined by
\begin{align}
\rho^*(\varepsilon) &=
\frac1{L}\sum_{\bm k}\delta\left[\varepsilon-\xi({\bm k})\right]
= \tilde{\phi}_\gamma \rho^{\rm (c)}(\varepsilon) ,
\end{align}
which is well defined even in the Heisenberg limit.
%
Since $4\tilde{W}_s^2/\tilde{\chi}_s(0)^2 = 4^2(k_{\rm B}T_{\rm K})^2$,
it follows that
\begin{align}
\frac1{4}J_Q(0,{\bm q}) \simeq  4(k_{\rm B}T_{\rm K})^2
\left[P(0,{\bm q})-P_0(0)\right]. 
\end{align}
%
The subtraction term is given by
\begin{align}
P_0(0) &=
2 \int \hskip-3pt d\varepsilon_1 \hskip-2pt \int \hskip-3pt\varepsilon_2
\hskip2pt \rho^*(\varepsilon_1)\rho^*(\varepsilon_2)
\frac{f(\varepsilon_1)-f(\varepsilon_2)}{\varepsilon_2-\varepsilon_1}.
\end{align}
The $T$ dependence of $P_0(0)$ is small.

The homogeneous $({\bm q}=0)$ and staggered $({\bm q}={\bm Q}_{L})$ components of $P(0,{\bm q})$ are given by
\begin{align}\label{EqJQ0}
P(0,0) &=  
2\int \hskip-3pt d\varepsilon\rho^*(\varepsilon)
\left[- \frac{d f(\varepsilon)}{d\varepsilon}
\right]  ,
\end{align}
and
\begin{align}\label{EqJQQ}
P(0,{\bm Q}_{L}) &= 
\int \hskip-3pt d\varepsilon \rho^*(\varepsilon) 
\frac1{\varepsilon}\tanh\left(\beta\varepsilon\right).
\end{align}
When $D\ne 2$, $\rho^*(\varepsilon)$ is almost constant around $\varepsilon\simeq 0$, as shown in Eq.~(\ref{EqFLR1}), 
it follows that
\begin{align}
\label{EqJQ0T1}
P(0,0) &\simeq 
2\rho^*(0)\left[1 +O(T^2) \right],
\end{align}
and
\begin{align}\label{EqJQQT1}
P(0,{\bm Q}_{L}) &\simeq 
2\rho^*(0)\left[
\ln \bigl(T_{\rm K}/T\bigr) + O(T^0)\right],
\end{align}
When $D=2$, $\rho^*(\varepsilon)$ is logarithmically diverging as $\varepsilon\rightarrow 0$, as shown in Eq.~(\ref{EqFLR1}), it follows that 
\begin{align}\label{EqJQ0T2}
P(0,0) &\simeq 
\frac{2\alpha_2}{|t_d^*|}\left[\ln |\tilde{\phi}_\gamma t_d^*/(k_{\rm B}T)| +O(T^0)\right],
\end{align}
and
\begin{align}\label{EqJQQT2}
P(0,{\bm Q}_{L}) &\simeq 
\frac{2\alpha_2}{|t_d^*|}\left\{ 
\bigl[\ln |\tilde{\phi}_\gamma t_d^*/(k_{\rm B}T)|\bigr]^2 + O(T^0)\right\}.
\end{align}
The $T$ dependence of $P(0,{\bm Q}_{L})$ or $J_Q(0,{\bm Q}_{L})$ is larger than that of $P(0,0)$ or $J_Q(0,0)$ because of the perfect nesting of the Fermi surface.
The strength of $J_Q(0,{\bm q})$ is proportional to the bandwidth of $\xi({\bm k})$ or  $k_{\rm B}T_{\rm K}$.
\cite{comItinerantMag} 

If either of $\delta\Sigma_{\sigma}({\rm i}\varepsilon_n,{\bm k})$ and $\Lambda(0,{\bm q})$ can be ignored, 
the staggered susceptibility shows a logarithmic $T$ dependence due to the perfect nesting of the Fermi surface, i.e., it approximately obeys the CW law; this is a mechanism of the CW law in antiferromagnetic metals.\cite{cwAF,FJO-Landau-AFCW}
Since the density of state $\rho^*(\varepsilon)$ has a sharp peak at $\varepsilon=0$ in two dimensions, the homogeneous susceptibility approximately obeys the CW law; this is a mechanism of the CW law in ferromagnetic metals.\cite{miyai}


In general, the ${\bm q}$ dependence of $\Lambda(0,{\bm q})$ is small.
Even if $\delta\Sigma_\sigma(\varepsilon+{\rm i}0,{\bm k})$ is considered, the Fermi surface shows a perfect nesting for ${\bm Q}_{L}$,
so that the temperature dependence of $J_Q(0,{\bm Q}_{L})$ must be much stronger than that of $J_Q(0,0)$. Thus,
the qualitative features of $\chi_s(0,{\bm Q}_{L})$ and $\chi_s(0,0)$
when $\delta\Sigma_\sigma(\varepsilon+{\rm i}0,{\bm k})$ is considered must be the same as those when $\delta\Sigma_\sigma(\varepsilon+{\rm i}0,{\bm k})$ is ignored; 
the $T$ dependence of $\chi_s(0,{\bm Q}_{L})$ is stronger than that of $\chi_s(0,0)$. Such ${\bm q}$ dependence is characteristic of itinerant-electron magnetism, and it can be distinguished from the ${\bm q}$ dependence of local-moment magnetism, where only the Weiss constant depends on ${\bm q}$.

\subsubsection{Magnetization in the N\'{e}el state}
When we follow previous papers,\cite{FJO-Landau-AFCW,FJO-Landau-MultiQ} it is straightforward to derive Landau's free energy below $T_{\rm N}$.  In two dimensions and higher, provided that critical fluctuations are ignored, staggered magnetization at $T\le T_{\rm N}$ and $T\simeq T_{\rm N}$ is given by
\begin{align}
|{\bm m}_{{\bm Q}_L}| \propto 
\sqrt{1-(T/T_{\rm N})^2},
\end{align}
either when $T_{\rm N} \ll T_{\rm K}$ or when $T_{\rm N} \gg T_{\rm K}$.

When $T_{\rm N} \gg T_{\rm K}$, an antiferromagnetic (AF) gap much larger than $k_{\rm B}T_{\rm K}$ opens at $T\ll T_{\rm N}$. The spectrum of single-particle excitations is not well defined at any $T$.
If an AF gap is smaller than $k_{\rm B}T_{\rm K}$, which is only possible when $T_{\rm N} \ll T_{\rm K}$, the spectrum of single-particle excitations is still well defined at $T\ll T_{\rm K}$, even at $T\le T_{\rm N}$, in the Heisenberg limit.

\section{Discussion}
\label{SecDiscussion}
The Kondo temperature $T_{\rm K}$ or $k_{\rm B}T_{\rm K}$ is the energy scale of local quantum spin fluctuations. When $T\gg T_{\rm K}$, local thermal spin fluctuations overcome the quantum ones so that
electron behave as localized spins. When $T_{\rm N}\gg T_{\rm K}$, therefore, magnetism is local-moment magnetism.
When $T\ll T_{\rm K}$, on the other hand,
electrons are itinerant even in the Heisenberg limit, in general.
An antiferromagnetic gap opens at $T\le T_{\rm N}$. 
Since the gap is a complete gap in the model of this paper, the conductivity vanishes at $T=0$~K, i.e., electrons are localized at $T=0$~K. 
If the gap is much smaller than $k_{\rm B}T_{\rm K}$, however,
electrons can still behave as itinerant ones so that 
magnetism is itinerant-electron magnetism.
If the gap is much larger than $k_{\rm B}T_{\rm K}$, which is presumably only possible when $T_{\rm N} \gg T_{\rm K}$, 
electrons cannot behave as itinerant ones so that magnetism is local-moment magnetism.
Whether magnetism is itinerant-electron or local-moment magnetism is determined by whether $T_{\rm N}\ll T_{\rm K}$ or $T_{\rm N} \gg T_{\rm K}$. 

It is an interesting issue whether or not the above discussion on the Heisenberg limit can be extended to the Heisenberg model.
The relationship between the Heisenberg and Hubbard models is similar to that between the $s$-$d$ and Anderson models.
According to Yosida's theory,\cite{yosida} the ground state of the $s$-$d$ model is a singlet.
Anderson's scaling theory supports the singlet ground state.\cite{poorman}
Wilson proved that the ground state is a singlet; a crossover occurs between a localized spin at $T\gg T_{\rm K}$ and a local spin liquid at $T \ll T_{\rm K}$.\cite{wilson}
Nozi\`{e}res proposed a Fermi-liquid description of the local spin liquid
based on the phase shift of conduction electrons due to scattering by the local spin liquid.\cite{nozieres} 
Either of the ground states in the $s$-$d$ and Anderson models is a singlet or a normal Fermi liquid.
Then, essentially the same Fermi-liquid analysis as Nozi\`{e}res's was made based on the Anderson model by Yamada and Yosida.\cite{yamada1,yamada2}
In the Heisenberg limit of the Hubbard model, 
the ground state within the constrained Hilbert subspace is also a singlet, a normal Fermi liquid or a Tomonaga-Luttinger (TL) liquid.
The Green function for the reservoir is given by
\begin{align}
G_{c\sigma}(\varepsilon+{\rm i}0,{\bm k}) &=
1/\left[\varepsilon + 2t_c\varphi({\bm k})
- \Gamma_c(\varepsilon + {\rm i}0)\right].
\end{align}
%
When $|\varepsilon|\ll k_{\rm B}T_{\rm K}$, it follows that
%
\begin{align}
\frac{\Gamma_c(\varepsilon \hskip-1pt + \hskip-1pt{\rm i}0)}{\lambda^2} &=
\frac{\overline{|v|^2}}{\tilde{\phi}_\gamma} 
\frac1{L}\! \sum_{\bm k} \frac1{\varepsilon \hskip-1pt - \hskip-1pt\xi({\bm k}) 
\hskip-1pt - \hskip-1pt \delta\Sigma_\sigma(\varepsilon \hskip-1pt + \hskip-1pt {\rm i}0, {\bm k})/ \tilde{\phi}_\gamma}.
\end{align}
Since $\overline{|v|^2}/\tilde{\phi}_\gamma =O(U^0)$,
$\Gamma_c(\varepsilon + {\rm i}0)/\lambda^2$ is well defined even in the Heisenberg limit. 
The quantity $\Gamma_c(\varepsilon + {\rm i}0)$ 
depends on vanishing fermionic excitations in the Hubbard model, and it corresponds to the phase shift discussed by Nozi\`{e}res.\cite{nozieres} 
A vanishing fermionic spectrum exists in the $s$-$d$ limit of the Anderson model or the Heisenberg limit of the Hubbard model.
It is reasonable that there is {\it a trace or vestige} of the vanishing fermionic spectrum even in the conduction band of the $s$-$d$ model or the reservoir for the Heisenberg model.

The charge susceptibility is exactly zero in the $s$-$d$ model but is nonzero in the Anderson model.
The conductivity is exactly zero in the Heisenberg model but is nonzero, finite or infinite, in the Hubbard model.
These differences are obvious because the local electron number at each unit cell is a conserved quantity in the $s$-$d$ and Heisenberg models but is not in the Anderson and Hubbard models, i.e., local gauge symmetry exists in the $s$-$d$ and Heisenberg models but it does not in the Anderson and Hubbard models.
On the other hand,
local gauge symmetry can never be spontaneously broken {\it or recovered}.\cite{elitzur}
The {\it recovery} of local gauge symmetry in the $s$-$d$ or Heisenberg model is caused by {\it a non-spontaneous process}, i.e. by constraining the Hilbert space within the subspace where no empty or double occupancy is allowed at a localized spin site or each unit call.
Because of this peculiar nature of local gauge symmetry, it is never a relevant symmetry to classify or distinguish a phase from the other. Thus, the adiabatic continuation\cite{adiabaticCont} holds between a spin liquid in the $s$-$d$ model and an electron liquid in the Anderson model, i.e., low-energy physical properties of a spin liquid in the $s$-$d$ model can be described as those of a normal Fermi liquid in the Anderson model.
According to the scaling theory by Abrahams et al.,\cite{abrahams}
there is no mobility edge between itinerant and localized states in a disordered system or there is no minimum metallic conductivity,
which means that there is no qualitative difference between a metal and an insulator.
Thus, the adiabatic continuation can hold between a metal and an insulator, in general; e.g., it holds between Wilson's insulator and a doped metal.
The adiabatic continuation must also hold between an electron liquid in the Hubbard model and a spin liquid in the Heisenberg model such that
low-energy physical properties of a spin liquid in the Heisenberg model can be described as those of an electron liquid in the Hubbard model.
Thus, it is reasonable to speculate that {\it itinerant-electron type of magnetism} is also possible in the Heisenberg model.
Since $T_{\rm K} \simeq T_{\rm MF}/(2D)$, where $T_{\rm MF}$ is the N\'{e}el temperature in the mean-field approximation for the corresponding Heisenberg model and $D$ is the spatial dimensionality,
whether magnetism is {\it itinerant-electron type of magnetism} or local-moment magnetism must be determined by  whether $T_{\rm N}\ll T_{\rm MF}/(2D)$ or $T_{\rm N} \gg T_{\rm MF}/(2D)$.

It has been proposed that, in one dimension, low-energy spin excitations in the Heisenberg and XXY models can be mapped to those of a TL liquid.\cite{TL1,TL2,TL3,TL4} 
This proposal is simply a proposal that magnetism in the Heisenberg and XXY models must be itinerant-electron type of magnetism.
%
According to the Kondo-lattice theory,
the spectrum of low-energy pair excitations in the Heisenberg limit are determined by the imaginary part of the exchange interaction $J_Q(\omega+{\rm i}0, {\bm q})$ or $P(\omega+{\rm i}0, {\bm q})$.
When the RVB term is considered but $\delta\Sigma_\sigma(\varepsilon + {\rm i}0, {\bm q})$ is ignored,
the energy $\omega_{\rm pair}(q)$ of a pair excitation lies in 
\begin{align}\label{EqDeCP0}
6\hskip1pt\Xi_1 |J| \sin(qa) \le \omega_{\rm pair}(q) 
\le 12\hskip1pt\Xi_1|J|\sin(qa/2),
\end{align}
as a function of wave number $q$.
The lower limit corresponds to des Cloizeaux-Pearson's mode in the Heisenberg model,\cite{cloiseaux}
whose dispersion relation is given by 
\begin{align}
\omega_{\rm dCP}(q) = \bigl(2|J|/\pi\bigr)\sin(qa).
\end{align}
When $\delta\Sigma_\sigma(\varepsilon + {\rm i}0, {\bm q})$ is ignored, $\Xi_{1}=1/\pi$, as shown in \S\ref{SecFS}. If $\Xi_{1}=1/\pi$, the lower limit is three times as large as $\omega_{\rm dCP}(q)$. 
The imaginary part of $\delta\Sigma_\sigma(\varepsilon + {\rm i}0, {\bm q})$ reduces $\Xi_D$.
The energy dependence of $\delta\Sigma_\sigma(\varepsilon+ {\rm i}0, {\bm q})$ also reduces $\Xi_D$. 
It is interesting to study how  $\omega_{\rm pair}(q)$ is corrected or modified when $\delta\Sigma_\sigma(\varepsilon\pm {\rm i}0, {\bm q})$ is considered.

In the Heisenberg model on the square lattice, the N\'{e}el state is only stabilized at $T=0$~K: $T_{\rm N}=+0$~K.
The staggered susceptibility $\chi_s(0,{\bm Q}_L)$ is diverging as $T\rightarrow 0$~K.
On the other hand, the homogeneous $\chi_s(0,0)$ shows a peak around $T_{\rm MF}\simeq |J|/k_{\rm B}$.\cite{Hz2D1,Hz2D2,Hz2D3}
This feature, $\chi_s(0,{\bm q})$ having qualitatively different $T$ dependences for different ${\bm q}$'s,
is characteristic of itinerant-electron magnetism.
For example, a similar result was also obtained by a conventional perturbative theory in terms of $U$ based on an itinerant electron model.\cite{miyake} 
When we follow previous papers,\cite{FJO-Landau-AFCW,miyai,miyake} it is easy to show that,  
since the Fermi surface shows a perfect nesting and the density of state $\rho^*(\varepsilon)$ has a sharp peak at the chemical potential,
$\Lambda(0,{\bm q})$ has a large $T$ dependence such that it suppresses the Curie-Weiss $T$ dependence for any ${\bm q}$; the ${\bm q}$ dependence of $\Lambda(0,{\bm q})$ is small.
The reduction of $T_{\rm N}$ and 
the peak around $T_{\rm MF}\simeq |J|/k_{\rm B}$ in $\chi_s(0,0)$ can also be explained in terms of the mode-mode coupling $\Lambda(0,{\bm q})$ by the Kondo-lattice theory based on the Hubbard model.
Furthermore, magnetization at $T=0$~K is as small as $0.3~\mu_{\rm B}$ per unit cell.\cite{reger} This small number implies that, in the corresponding Heisenberg limit of the Hubbard model, an antiferromagnetic gap is smaller than $k_{\rm B}T_{\rm K}$
so that magnetism is itinerant-electron magnetism.
Thus, we propose that, even in the Heisenberg model in two dimensions, 
magnetism at $T \ll T_{\rm MF}/4$ should be characterized as itinerant-electron type of magnetism.


\section{Conclusion}
\label{SecConclusion}
The Heisenberg limit of the Hubbard model is studied by the Kondo-lattice theory.
The Kondo temperature $T_{\rm K}$ or $k_{\rm B}T_{\rm K}$
can be still defined as an energy scale of low-energy local quantum spin fluctuations even in the Heisenberg limit. 
It is enhanced by 
the resonating valence bond (RVB) mechanism, so that $T_{\rm K} \simeq T_{\rm MF}/(2D)$, where $T_{\rm MF}$ is the N\'{e}el temperature in the mean-field approximation of the corresponding Heisenberg model and $D$ is the spatial dimensionality. 
When $T \gg T_{\rm K}$, electrons behaves as localized spins and the entropy is about $k_{\rm B}\ln 2$ per unit cell. Thus, a high temperature phase at $T\gg T_{\rm K}$ is a Mott insulator. 
Within the constrained Hilbert subspace where no order parameter exists, however, the ground state is not a Mott insulator nor a Lieb-Wu insulator either
but is an almost spin liquid, i.e., a normal Fermi liquid or a Tomonaga-Luttinger liquid in which
the band of low-energy single-particle excitations is almost vanishing; however, the width of the vanishing band is $O(k_{\rm B}T_{\rm K})$, the Fermi surface is well defined in the vanishing band, and the conductivity at $T=0$~K is diverging in the vanishing limit of disorder. 

When $D\gg1$, the N\'{e}el temperature $T_{\rm N}$ is so high that $T_{\rm N}\simeq T_{\rm MF}$ but the Kondo temperature is so low that $T_{\rm K}\ll T_{\rm MF}$. Thus, $T_{\rm N}\gg T_{\rm K}$,
so that magnetism for $D\gg1$ is local-moment magnetism.
In particular, magnetism in infinite dimensions, where $T_{\rm K}=+0$\hskip2ptK and $T_{\rm N}=T_{\rm MF}$, is prototypic local-moment magnetism.
When $D$ is small enough, $T_{\rm K}$ is so high or
$T_{\rm N}$ is so low that $T_{\rm N}\ll T_{\rm K}$.
Electrons are itinerant at $T\ll T_{\rm K}$,
unless an antiferromagnetic complete gap opens in the vanishing band.  
Magnetic properties at $T_{\rm N}<T\ll T_{\rm K}$ and $|\omega|\ll k_{\rm B}T_{\rm K}$ are those of a normal Fermi liquid or a Tomonaga-Luttinger liquid, i.e., an itinerant electron liquid; e.g., the spin susceptibility has a temperature and wave-number dependence characteristic of itinerant-electron magnetism. 
When $T_{\rm N}\ll T_{\rm K}$, therefore, magnetism is itinerant-electron magnetism.

Since local gauge symmetry is never relevant to classify a phase from the other,
there is no essential difference between an electron liquid in the Hubbard model, where local gauge symmetry does not exist, and a spin liquid in the Heisenberg model, where local gauge symmetry exists, except for charge-channel properties such as the conductivity.  Thus, {\it itinerant-electron type of magnetism} must also be possible even in the Heisenberg model.
Whether magnetism is itinerant-electron type of magnetism or local-moment magnetism must be determined by whether $T_{\rm N}\ll T_{\rm MF}/(2D)$ or $T_{\rm N} \gg T_{\rm MF}/(2D)$.
Since no $T_{\rm N}$ exists in one dimension and $T_{\rm N}=+0$~K in two dimensions, in particular, 
magnetic properties of the Heisenberg model at $T \ll T_{\rm MF}$ and $|\omega|\ll k_{\rm B}T_{\rm MF}$ in one dimension and two dimensions must be able to be described as those of an itinerant-electron liquid, i.e.,
a Tomonaga-Luttinger liquid or a normal Fermi liquid.

\section*{Acknowledgements}
The author thanks to T. Hikihara for useful discussions.

\appendix
\section{Proof of Eq.~(\ref{EqWardtdtd*})}
\label{SecEquality}
Consider a function defined by
\begin{align}\label{EqIntC1}
X(\mu^*) &=
\frac1{L}\sum_{\bm k} \varphi({\bm k})
f\bigl[\xi({\bm k})-\mu^*\bigr] 
\nonumber \\ &=
\frac{\sqrt{D}}{L}\sum_{\bm k} \cos(k_1a) 
f\bigl[\xi({\bm k})-\mu^*\bigr] ,
\end{align}
where $\xi({\bm k})=-2\bigl(t_d^*/\tilde{\phi}_\gamma\bigr) \varphi({\bm k})$.
Equation (\ref{EqIntC1}) is also given, in the integration form, by
\begin{align}\label{EqIntC2}
X(\mu^*) &=
\frac{\sqrt{D}a^D}{(2\pi)^D} \hskip-4pt
\int_{-\pi/a}^{+\pi/a} \hskip-18pt dk_1 \cdots \hskip-3pt 
\int_{-\pi/a}^{+\pi/a} \hskip-18pt dk_D  \hskip0pt
\cos(k_1a) 
f\bigl[\xi({\bm k})\hskip-1pt - \hskip-1pt\mu^*\bigr].
\end{align}
By partial integration of Eq.~(\ref{EqIntC2}) with respect to $k_1$, it follows that 
\begin{align}\label{EqC=S}
X(\mu^*)= 2\left(t_d^*/\tilde{\phi}_\gamma\right) Y(\mu^*), 
\end{align}
where
\begin{align}\label{EqIntS1}
Y(\mu^*) &= 
\frac{a^D}{(2\pi)^D} \hskip-2pt\int_{-\pi/a}^{+\pi/a}  \hskip-10pt dk_1 \cdots \hskip-2pt\int_{-\pi/a}^{+\pi/a} \hskip-10pt dk_D\hskip2pt
\hskip-2pt\sin^2(k_1a)\hskip-2pt 
\nonumber \\ & \qquad \times 
\left[- \frac{df(\varepsilon)}{d\varepsilon}\right]_{\varepsilon=\xi({\bm k})-\mu^*}.
\end{align}
This is also given, in the sum form, by
\begin{align}\label{EqIntS2}
Y(\mu^*) &=
\frac{1}{L}\sum_{\bm k} \sin^2(k_1a)
\left[- \frac{df(\varepsilon)}{d\varepsilon}\right]_{\varepsilon=\xi({\bm k})-\mu^*}
\nonumber \\ &=
\frac{1}{L}\sum_{\bm k} \frac1{D}\sum_{\nu=1}^{D}\sin^2(k_\nu a)
\left[- \frac{df(\varepsilon)}{d\varepsilon}\right]_{\varepsilon=\xi({\bm k})-\mu^*}.
\end{align}

Since $\Xi_{D}=X(0)$ and $\pi_{xx}(0) = Y(0)$, it follows from Eq.~(\ref{EqC=S}) that $\Xi_{D}=2(t_d^*/\tilde{\phi}_\gamma)\pi_{xx}(0)$ or
\begin{align}\label{EqXiPi}
\Xi_{D} &=
\frac{2}{\tilde{\phi}_\gamma}\left(t_d - \frac{3}{4}\tilde{\phi}_\gamma \tilde{W}_s^2 \Xi_{D}\frac{J}{D}\right)
\pi_{xx}(0).
\end{align}
Then, it is straightforward to prove Eq.~(\ref{EqWardtdtd*}).



\begin{thebibliography}{99}
\bibitem{comFrustration}
The specific heat linear in $T$ at low temperatures is due to the accumulation of low-energy excitations.
Such accumulation in a system is only possible if the system is so frustrated that no order parameter exists.
In this sense, a normal Fermi liquid is frustrated, even if it is a free electron liquid.
If an order parameter exists or a system is not frustrated,
either of the specific heat and the entropy approaches zero 
faster than linearly in $T$  as $T\rightarrow0$~K, in general.
\bibitem{comDMFT1}
It is easy to prove that, when $T=0$~K, $\rho(0)=1/[\pi\Delta(0)]$ does not depend on $U/|t_d|$ in the S$^3$A. If the RVB mechanism is considered, $\rho(0)|t_d|\rightarrow +0$ as $D\rightarrow +\infty$ and $U/|t_d|\rightarrow +\infty$. Thus, the ground state in the S$^3$A is slightly different from that in the vanishing limit of the RVB mechanism or in the large limit of $D$ although either of the ground states is a normal Fermi liquid characterized by $T_{\rm K}=+0$~K.
\bibitem{comDMFT2}
Since the RVB mechanism is not considered in the S$^3$A, the probability $p$ of empty or double occupancy in the S$^3$A must be much smaller than $t_d^2/(DU^2)$. When $U/|t_d|\gg 1$,  $p$ may depend on $\lambda^2$, i.e., $p=O(\lambda^2)$. When $U/|t_d|\gg 1$,
a solution of a Mott insulator was obtained in numerical theories based on DMFT,\cite{PhyToday,RevMod,kotliar} in which $p$ is presumably vanishingly small or exactly zero. The solution implies that, in the S$^3$A or DMFT, $p=O(\lambda^2)$ if the electron reservoir is explicitly considered.
\bibitem{comItinerantMag} 
If two limiting procedures of $T\rightarrow 0$\hskip2ptK and $D\rightarrow+\infty$ are taken so as to keep $T/T_{\rm K}\ll 1$, $J_Q(0,{\bm q})$ is $O(k_{\rm B}T_{\rm K}/D^0)$ for particular ${\bm q}$'s although it is $O\bigl(k_{\rm B}T_{\rm K}/\sqrt{D}\bigr)$ for almost all ${\bm q}$'s.
%
When $t_d$ is finite, the electron filling is non-half filling, the nesting of the Fermi surface is not sharp, and the superexchange interaction is weak enough, i.e., in a non-Heisenberg-limit case, it is possible that $T_{\rm N}\ll T_{\rm K}$ even in infinite dimensions. 
When $T_{\rm N}\ll T_{\rm K}$ in infinite dimensions,
magnetism is prototypic itinerant-electron magnetism.
\end{thebibliography}
\end{document}